\documentclass[%
 reprint,
 amsmath,amssymb,
 aps,
]{revtex4-2}
\usepackage{physics}
\usepackage[hidelinks]{hyperref}
\usepackage{xcolor}
\usepackage{pgf}   % Required for PGF files
\usepackage{bm}% bold math

\hypersetup{
    citecolor=blue,
    filecolor=blue,
    linkcolor=blue,
    urlcolor=black
}

\newcommand{\ch}{\hat{c}}

\usepackage{ifthen}
\newboolean{showcomments}
\setboolean{showcomments}{true}

\begin{document}

%\preprint{APS/123-QED}

%\title{Finite-island corrections to the charge qubit spectrum and susceptibility}
\title{Quantum theory of the Josephson junction between finite islands}
\author{Thomas J. Maldonado}
\email{maldonado@princeton.edu}
\affiliation{%
Department of Electrical and Computer Engineering, Princeton University, Princeton, NJ 08544, USA}
\author{Alejandro W. Rodriguez}%
\affiliation{%
Department of Electrical and Computer Engineering, Princeton University, Princeton, NJ 08544, USA}
\author{Hakan E. Türeci}%
\affiliation{%
Department of Electrical and Computer Engineering, Princeton University, Princeton, NJ 08544, USA}

\date{\today}% It is always \today, today,

\begin{abstract}
    Superconducting circuits comprising Josephson junctions have spurred significant research activity due to their promise to realize scalable quantum computers. Effective Hamiltonians for these systems have traditionally been derived assuming the junction connects superconducting islands of infinite size. We derive a quantized Hamiltonian for a Josephson junction between finite-sized islands and predict measurable corrections to the qubit frequency and charge susceptibility to test the theory.
    
\end{abstract}

\maketitle
In an isolated system of two weakly linked superconducting islands, the maximum achievable charge imbalance is determined by the total number of superconducting charge carriers. In contrast, prevailing quantum descriptions of these circuit elements~\cite{blais2021circuit} permit unbounded charge imbalances, an approximation valid only in the limit of large superconducting islands and negligible gate voltage. The opposite regime has remained relatively unexplored and constitutes the focus of this Letter. Here, we present a quantum theory of tunneling between finite superconducting islands and use this theory to predict observable departures from the typical infinite-island theories employed in the literature.

We draw our inspiration from the Bose-Einstein condensate (BEC)~\cite{griffin1996bose}---a state of matter embodied by a macroscopic wave function common to a collection of bosons at low temperatures---which has been experimentally observed in a variety of ultracold atomic gases~\cite{anderson1995observation, davis1995bose}. A striking parallel to trapped BECs is found near a superconductor's critical temperature, where the bosonic quasiparticles known as Cooper pairs~\cite{BCS} exhibit an effective description via the Ginzburg-Landau order parameter~\cite{GL} that is functionally analogous to the BEC's macroscopic wave function. In contrast to the Gross-Pitaevskii equation typically used to describe the evolution of BECs~\cite{bao2003numerical}, the Ginzburg-Landau equations are only equipped to model the order parameter's steady state. Nonetheless, far below the critical temperature, two superconductors separated by a thin insulating region~\cite{josephson1962possible, josephson2} exhibit dynamics so similar to a BEC in a double well potential~\cite{gati2007bosonic} that both systems bear the same name: the Josephson junction (JJ)---to distinguish the two, the latter is sometimes called a bosonic JJ.

The literature contains semiclassical studies of superconducting~\cite{maldonado2025mesoscopic} and bosonic~\cite{raghavan1999coherent} JJs that suggest comparable finite-size effects, but a corresponding analysis at the quantum level remains underdeveloped. To address this gap, we model the quantum dynamics of a JJ between finite superconducting islands via the two-site Bose-Hubbard Hamiltonian
\begin{align}\label{eq:BH}
        \hat{H}  &= \lambda\left(\ch^\dagger_{\mathrm{L}}\ch^\dagger_{\mathrm{L}}\ch^{}_{\mathrm{L}}\ch^{}_{\mathrm{L}} + \ch^\dagger_{\mathrm{R}}\ch^\dagger_{\mathrm{R}}\ch^{}_{\mathrm{R}}\ch^{}_{\mathrm{R}}\right) + \mu\left(\ch^\dagger_\mathrm{L}\ch^{}_\mathrm{L}-\ch^\dagger_\mathrm{R}\ch^{}_\mathrm{R}\right) \nonumber\\ 
        & \quad -\nu\left(\ch_{\mathrm{L}}^\dagger \ch^{}_{\mathrm{R}} + \ch_{\mathrm{R}}^\dagger \ch^{}_{\mathrm{L}}\right),
\end{align}
in direct analogy with the bosonic JJ~\cite{gati2007bosonic, links2006two, PhysRevA.55.4318, leggett2001bose}. 
Here, \(\hat{c}_{\mathrm{L}}\) and \(\hat{c}_{\mathrm{R}}\) denote the annihilation operators for the superconducting bosons (i.e., Cooper pairs) to the left and right of the JJ, which satisfy \([\hat{c}^{}_\mathrm{L},\hat{c}_\mathrm{L}^\dagger]=[\hat{c}^{}_\mathrm{R},\hat{c}_\mathrm{R}^\dagger]=1\), the only nontrivial commutation relations. As usual, the free parameters \(\lambda\), \(\mu\), and \(\nu\) represent the on-site interaction strength, (twice) the voltage-induced chemical potential bias, and the tunneling amplitude, respectively. 
We note that while Eq.~\eqref{eq:BH} has predominantly been employed in the context of trapped BECs, it has also seen limited application in superconducting systems~\cite{benatti2008cooper, alicki2008charge, dell2012analytical}, though a systematic treatment of its finite-size implications for the observables central to circuit quantum electrodynamics (cQED)~\cite{blais2021circuit} is lacking. We proceed by reformulating Eq.~\eqref{eq:BH} in terms of the spin operators
\begin{subequations}
    \begin{align}
        \hat{S}_x &= \frac{1}{2}\left(\hat{c}^\dagger_\mathrm{L}\hat{c}^{}_\mathrm{R}+\hat{c}^\dagger_\mathrm{R}\hat{c}^{}_\mathrm{L}\right),\\
        \hat{S}_y &= \frac{1}{2i}\left(\hat{c}^\dagger_\mathrm{L}\hat{c}^{}_\mathrm{R} - \hat{c}^\dagger_\mathrm{R}\hat{c}^{}_\mathrm{L}\right),\\
        \hat{S}_z &= \frac{1}{2}\left(\hat{c}^\dagger_{\mathrm{L}}\hat{c}^{}_{\mathrm{L}} - \hat{c}^\dagger_{\mathrm{R}}\hat{c}^{}_{\mathrm{R}}\right),
    \end{align}
\end{subequations}
which satisfy the SU(2) commutation relations \([\hat{S}_j,\hat{S}_k]=i\epsilon_{jkl}\hat{S}_l\) and \(\hat{S}_x^2 + \hat{S}_y^2 + \hat{S}_z^2 = N(N+1)\)~\cite{osti_4389568}. Here, we have fixed the total number of superconducting bosons per island \(N \equiv (\ch^\dagger_\mathrm{L}\ch^{}_\mathrm{L} + \ch^\dagger_\mathrm{R}\ch^{}_\mathrm{R})/2 \in \mathbb{N}/2\), since it is a conserved quantity \([N,\hat{H}]=0\). In terms of the cQED parameters \(E_\mathrm{J} \equiv 2N\nu>0\), \(E_\mathrm{C} \equiv 2\lambda>0\), and \(n_\mathrm{g} \equiv -\mu/(2\lambda)\), Eq.~\eqref{eq:BH} now reads
\begin{align}\label{eq:quantum_H}
    %\begin{aligned}
        \hat{H}
        &= E_\mathrm{C} \left(\hat{S}_z-n_\mathrm{g}\right)^2 - \frac{E_\mathrm{J}}{N}\hat{S}_x\nonumber\\
        &= \sum_{n = -N}^N E_\text{C}\left(n-n_\text{g}\right)^2\ket{n}\bra{n}\nonumber
        \\
        &-\sum_{n=-N}^{N-1}\frac{E_\text{J}}{2N}\sqrt{N(N+1)-n(n+1)}\left(\ket{n}\bra{n+1} + \mathrm{H.c.}\right)
    %\end{aligned}
\end{align}
up to a constant shift in energy with H.c. short for Hermitian conjugate. In the second equality, we have introduced the finite orthonormal basis of eigenstates satisfying \(\hat{n}\ket{n} = n\ket{n}\) with \(\hat{n} = \hat{S}_z\) the number operator. The eigenvalues are given by \(n\in\{-N,...,N\}\) with consecutive elements differing by 1.
\begin{figure}
    %\hspace*{-1cm}
    \centering
    \includegraphics{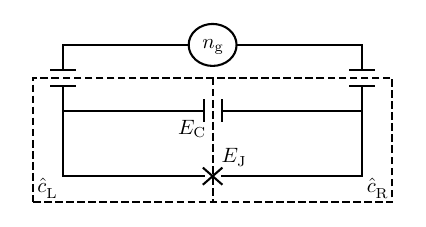}  % Directly embed the PGF figure
    \caption{The circuit diagram depicts two superconducting islands (dashed lines) separated by a Josephson junction with Josephson energy \(E_\text{J}\) and capacitive charging energy \(E_\text{C}\). A voltage proportional to the (dimensionless) offset charge \(n_\text{g}\) is applied via capacitive coupling, where the constant of proportionality and all other device parameters are related to the geometric and material properties of the islands via the predictions in~\cite{maldonado2025mesoscopic}. We formulate a quantum theory for the population imbalance \(\hat{n} = (\ch^{\dagger}_\mathrm{L}\ch^{}_\mathrm{L} - \ch^{\dagger}_\mathrm{R}\ch^{}_\mathrm{R} )/2\) in a circuit with \(N = (\ch^{\dagger}_\mathrm{L}\ch^{}_\mathrm{L} + \ch^{\dagger}_\mathrm{R}\ch^{}_\mathrm{R} )/2\) bosons per island, where \(\ch^{}_\text{L}\) and \(\ch^{}_\text{R}\) denote the annihilation operators for bosons in the left and right islands.}
    \label{fig:circuit}
\end{figure}
\begin{figure*} % Use figure* to span across both columns
    \hspace*{-0.75cm}
    \centering
    \includegraphics{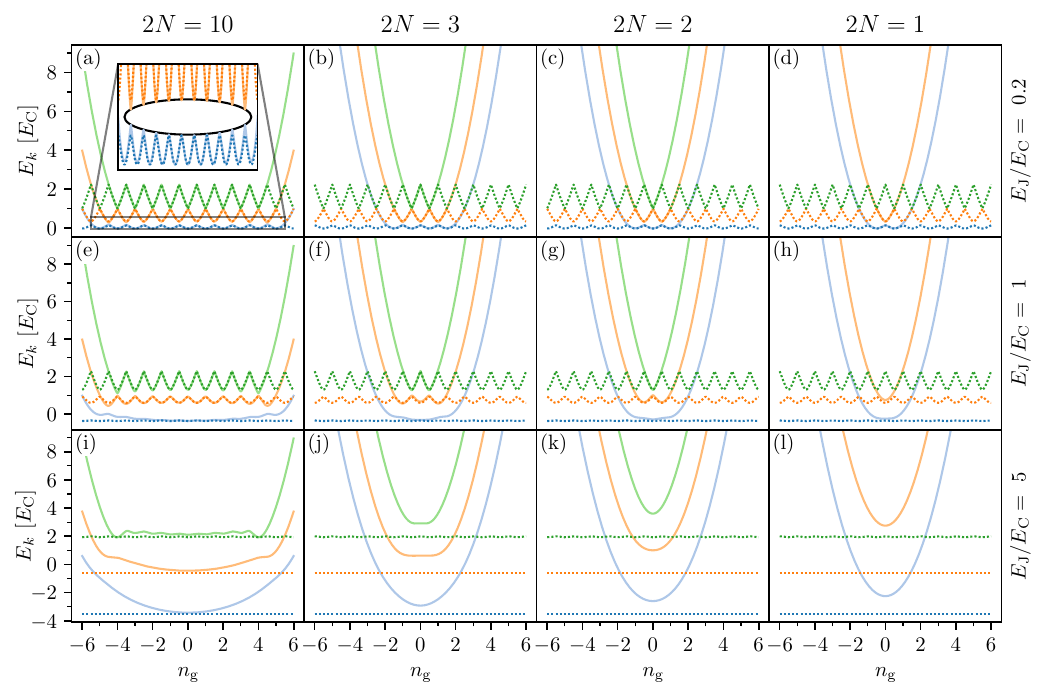}
\caption{Energy eigenvalues \(E_k\) (first three levels, \(k=0,1,2\)) of the finite-island Hamiltonian [Eq.~\eqref{eq:quantum_H}] vs. offset charge \(n_\text{g}\) are plotted as solid colored curves for a range of \(E_\text{J}/E_\text{C}\) ratios (labels to the right of each row) and total number of bosons \(2N\) (labels at the top of each column). The \(N\rightarrow \infty\) limit of each band is plotted as a dotted curve of matching color. For \(\abs{n_\text{g}}\ll N\), the infinite-island spectrum is recovered. For \(\abs{n_\text{g}}\gg N\), anharmonicity is lost, since \(E_k \approx n_\text{g}^2 + 2\abs{n_\text{g}}k\) asymptotically. Changing \(2N\) by \(\pm 1\) yields parity switches: half-integer shifts in the local extrema of each band [e.g., compare (b) and (c)]. For \(E_\text{J}/E_\text{C}\ll 1\) [see (a)-(d)], the first band gap [Eq.~\eqref{eq:qubit_frequency}] takes the form of an ellipse [solid black curve of energy \(E_\text{C}/4 \pm \hbar\omega_\text{q}/2\) in the inset of (a)].}
    \label{fig:bands}
\end{figure*}

For reasonably sized islands \(N\gg 1\) with quantum states \(\ket{\psi}\) featuring a relatively small expected charge imbalance \(\bra{\psi}\hat{n}\ket{\psi} \ll N\), the off-diagonal matrix elements in Eq.~\eqref{eq:quantum_H} will be well-approximated by \(-E_\mathrm{J}/2\), and the dynamics \(i\hbar\dot{\ket{\psi}} = \hat{H}\ket{\psi}\) will be effectively described by the charge qubit Hamiltonian \(\hat{H} \approx E_\mathrm{C}(\hat{n}-n_\mathrm{g})^2-E_\mathrm{J}\cos\hat{\varphi}\) with \([\hat{\varphi},\hat{n}]=i\). In the appropriate limit, Eq.~\eqref{eq:quantum_H} thus reproduces the standard cQED description of a JJ~\cite{blais2021circuit}---often summarized by a lumped-element model such as that shown in Fig.~1---while still marking a clear departure for finite islands. If true, the proposed correspondence between JJs in superconductors and BECs would thus yield measurable corrections to standard observables like the qubit frequency and charge susceptibility. While experiments in the saturation regime \(n_\text{g} \sim N\) are uncommon, spectroscopic probes and single-electron transistors used to measure these quantities have already been well-developed by experimentalists in the field of quantum information~\cite{von2001spectroscopy, bladh2005single}. It is thus of fundamental interest to compute these corrections.

We first calculate the qubit frequency \(\omega_\text{q}\equiv (E_1 - E_0)/\hbar\), where \(E_k\) denotes the \(k\)th largest eigenvalue of \(\hat{H}\). The first three eigenvalues are plotted in Fig.~\ref{fig:bands} for a range of \(N\) and \(E_\text{J}/E_\text{C}\) ratios. The numerical results showcase convergence to the predictions from the Mathieu equation~\cite{cottet2002implementation} for \(\abs{n_\text{g}} \ll N\) and a loss of anharmonicity for \(\abs{n_\text{g}}\gg N\) (energy eigenvalues become equally spaced by \(\approx 2E_\text{C}\abs{n_\text{g}}\)). Extensive experimental efforts have been dedicated to the fabrication of qubits in the transmon regime \(E_\mathrm{J}/E_\mathrm{C}\gg 1\), where the reduced sensitivity of \(\omega_\mathrm{q}\) to \(n_\mathrm{g}\) (e.g., the flattened bands in Fig.~\ref{fig:bands}(i)) yields an increased resilience to charge noise. To this end, we calculate the qubit frequency for a standard transmon (\(E_\text{J}/\hbar = 10~\text{GHz}\) and \(E_\text{C}/\hbar = 0.2~\text{GHz}\)~\cite{PhysRevA.76.042319}) comprising \(1\times 10^9\) electrons (\(N= 10^{9}/4\) Cooper pairs per island) and report a frequency shift of \(\omega_\text{q}({n_\text{g} = \pm 10^6})-\omega_\text{q}(n_\text{g} =0) \approx -8~\text{kHz}\), which may be measured with standard spectroscopic probes developed for quantum computing~\cite{kristen2020amplitude} by applying a gate voltage of \(V_\mathrm{g} \equiv n_\mathrm{g}\cdot(2e/C_\mathrm{g})\approx 1~\text{kV}\) (assuming a gate capacitance of \(C_\mathrm{g}\approx 2e/\text{mV}\)~\cite{kringhoj2020suppressed}) to aluminum islands of volume \(\mathcal{V} \equiv N/n_\mathrm{s}\approx 0.005 ~\mu\text{m}^3\) [assuming a zero-temperature Cooper pair density \(n_\mathrm{s} \approx m_\mathrm{e}/(2\mu_0 e^2\lambda_\mathrm{L}^2)\) determined by the London penetration depth \(\lambda_\text{L}\approx 16~\text{nm}\)~\cite{kittel2018introduction}].
We note that this final assumption---a volume-independent Cooper pair density derived from the material's bulk properties---becomes unreliable for sufficiently small islands. In particular, superconductivity is expected to break down when the average level spacing of the Fermi sea \(\delta \approx \epsilon_\mathrm{F}/(n_\mathrm{e} \mathcal{V})\) exceeds the superconducting energy gap \(\Delta\)~\cite{von2001spectroscopy}; here, \(\epsilon_\mathrm{F}\) and \(n_\mathrm{e}\) denote the Fermi energy and electron density, respectively. We thus require \(\Delta \gtrsim \delta\) or equivalently
\begin{equation}
    N \gtrsim  \frac{\epsilon_\mathrm{F}}{\Delta}\frac{n_\mathrm{s}}{n_\mathrm{e}}.
\end{equation}
Using \(\Delta \approx 0.34\) meV, \(\epsilon_\mathrm{F} \approx 11.63\) eV, and \(n_\mathrm{e} \approx 18.06 \times 10^{22}~\mathrm{cm}^{-3}\) for aluminum~\cite{kittel2018introduction}, we find a lower bound of approximately \(N \gtrsim 1.0\times 10^4\), so the aforementioned parameters reside safely within the theory's regime of validity. Experimental challenges include the fabrication of small islands while maintaining standard values for \(E_\mathrm{J}\) and \(E_\mathrm{C}\), as well as the application of a large external voltage in a way that does not damage neighboring circuit elements (e.g., via a bias tee).

Despite the favorable \(\mathcal{O}(N)\) runtime of the above computation, representative analytical formulas are desirable. To this end, we treat the off-diagonal terms in Eq.~\eqref{eq:quantum_H} as a perturbation with \(E_\text{J}/E_\text{C} \ll 1\) the perturbative parameter and apply degenerate perturbation theory at the degeneracy points \(n_\text{g} \in \{-N + 1/2,...,N-1/2\}\). In the degenerate (low-energy) subspace \(\text{span}\{\ket{\lfloor n_\text{g}\rfloor}, \ket{\lceil n_\text{g}\rceil}\}\) with \(\lfloor n_\text{g} \rfloor \equiv \sup\{n \in \{-N,...,N\}~|~ n<n_\text{g}\}\) and \(\lceil n_\text{g} \rceil \equiv \inf\{n\in\{-N,...,N\}~|~n>n_\text{g}\}\), the Hamiltonian reads
\begin{equation}\label{eq:degenerate_H}
    \hat{H} \approx E_\text{C}\left(\hat{n}-n_\text{g}\right)^2  -\frac{E_\text{J}}{2N}\sqrt{N(N+1)-\lfloor n_\text{g}\rfloor \lceil n_\text{g}\rceil}\sigma_x
\end{equation}
with the effective number operator
\begin{equation}\label{eq:two_level_number}
    \hat{n} \approx \frac{\lfloor n_\text{g}\rfloor + \lceil n_\text{g} \rceil}{2} + \frac{\lfloor n_\text{g} \rfloor - \lceil n_\text{g}\rceil}{2}\sigma_z.
\end{equation}
Explicit diagonalization of Eq.~\eqref{eq:degenerate_H} yields two nondegenerate energy eigenvalues whose difference is given by
\begin{equation}\label{eq:qubit_frequency}
    \hbar\omega_\text{q} \approx \frac{E_\text{J}}{2N}\sqrt{\left(1+2N\right)^2-4n_\mathrm{g}^2},
\end{equation}
a good approximation at the aforementioned degeneracy points in a Cooper pair box \(E_\text{J}/E_\text{C} \ll 1\). The elliptic structure of this band gap is illustrated in Fig.~\ref{fig:bands}(a), and the expected result \(\lim_{N\rightarrow \infty} (\hbar\omega_\mathrm{q}) = E_\mathrm{J}\) is recovered in the infinite-island (i.e., thermodynamic) limit~\cite{bouchiat1998quantum}.

Analytical formulas for the transmon \(E_\text{J}/E_\text{C} \gg 1\)~\cite{PhysRevA.76.042319} are obtained via the Holstein-Primakoff transformation~\cite{PhysRev.58.1098, hirsch2013virtues}
\begin{subequations}\label{eq:HP}
    \begin{align}
    \hat{S}_x &= N-\hat{a}^\dagger \hat{a}\\
    \hat{S}_y &= \frac{1}{2}\left(\sqrt{2N-\hat{a}^\dagger\hat{a}}~\hat{a}+\mathrm{H.c.}\right)
    \\
    \hat{S}_z &= \frac{1}{2i}\left(\sqrt{2N-\hat{a}^\dagger\hat{a}}~\hat{a}-\mathrm{H.c.}\right)\label{eq:S_z}
    \end{align}
\end{subequations}
with the bosonic operators satisfying \([{\hat{a}},{\hat{a}^\dagger}] = 1\). In the low-energy subspace of the transmon, the expected bosonic excitation number remains much smaller than the dimension of the physical Hilbert space: \(E_\mathrm{J}/E_\mathrm{C} \gg 1 \implies \langle \hat{a}^\dagger \hat{a} \rangle \ll 2N+1\). To exploit this, we first apply a Taylor expansion to the square root term in Eq.~\eqref{eq:S_z}
\begin{equation}\label{eq:taylor}
    \sqrt{2N - \hat{a}^\dagger \hat{a}} =\sqrt{2N}\left(1 - \frac{\hat{a}^\dagger \hat{a}}{4N}+ {\mathcal{O}\left(\frac{\hat{a}^\dagger\hat{a}}{2N}\right)^2}\right)
\end{equation}
and partition the Hamiltonian \(\hat{H} = \hat{H}_0 + \delta \hat{H}\) into its leading order component \(\hat{H}_0\) and a correction term \(\delta \hat{H}\) that collects all higher-order contributions
\begin{subequations}\label{eq:partition}
\begin{align}
    \hat{H}_0 &= E_\mathrm{C} \left( \sqrt{N} \hat{p} - n_\mathrm{g} \right)^2-\frac{E_\mathrm{J}}{N}\left(N-\hat{a}^\dagger\hat{a}\right), \\
    \delta \hat{H} &= E_\mathrm{C} \left[ \left( \sqrt{N} \hat{p} - n_\mathrm{g} \right) \delta \hat{S}_z + \text{H.c.} \right] + E_\mathrm{C} \delta \hat{S}_z^2.
\end{align}
\end{subequations}
In the above equations, we have introduced the bosonic momentum operator \(\hat{p} \equiv i(\hat{a}^\dagger - \hat{a})/\sqrt{2}\) and the deviation of the spin \(z\) component from its leading-order form  \(\delta \hat{S}_z \equiv \hat{S}_z - \sqrt{N} \hat{p}\). We now treat \(\delta \hat{H}\) as a perturbation, since for sufficiently large \(E_\mathrm{J}/E_\mathrm{C}\), the total Hamiltonian \(\hat{H}\) will be well-approximated by \(\hat{H}_0\). The latter takes the form of a harmonic oscillator
\begin{equation}\label{eq:harmonic}
    \hat{H}_0 = \epsilon \hat{b}^\dagger\hat{b}
\end{equation}
 with level spacing \(\epsilon = \sqrt{2E_\mathrm{C}E_\mathrm{J} + E_\mathrm{J}^2/N^2}\). To arrive at Eq.~\eqref{eq:harmonic}, we have introduced the bosonic operators \([{\hat{b}},{\hat{b}^\dagger}]=1\) via the affine Bogoliubov transformation
\begin{subequations}
    \begin{align}
        \begin{pmatrix}
            \hat{b}\\\hat{b}^\dagger
        \end{pmatrix}
        &= \begin{pmatrix}
            u_+ & u_-\\
            u_- & u_+
        \end{pmatrix}
        \begin{pmatrix}
            \hat{a}
            \\
            \hat{a}^\dagger
        \end{pmatrix}
       +u_0\begin{pmatrix}
            -i\\+i
        \end{pmatrix}\\
        u_\pm &= \left(E_\mathrm{J} \pm N \epsilon\right)/\sqrt{4N\epsilon E_\mathrm{J}}.\\
        u_0&= n_\mathrm{g}\sqrt{2E_\mathrm{C}^2E_\mathrm{J}/\epsilon^3}
    \end{align}
\end{subequations}
and dropped the zero-point energy.
Before proceeding to the perturbative corrections, we note that the level spacing in the infinite-island limit  agrees with the well-known result \(\lim_{N\rightarrow \infty}\epsilon = \sqrt{2E_\mathrm{C}E_\mathrm{J}}\)~\cite{PhysRevA.76.042319}. We now apply first-order perturbation theory to approximate the qubit frequency
\begin{figure}
    \hspace*{-0.75cm}
    \centering
    \includegraphics{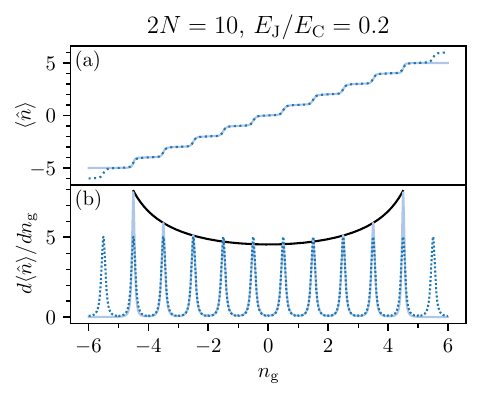}
    \caption{The expected population imbalance \(\langle \hat{n}\rangle = \bra{\psi_0(n_\text{g})}\hat{n}\ket{\psi_0(n_\text{g})}\) in the ground state \(\ket{\psi_0(n_\text{g})}\) of the finite-island Hamiltonian [Eq.~\eqref{eq:quantum_H}] vs. offset charge \(n_\text{g}\) is plotted as a solid blue curve in (a) with \(2N=10\), \(E_\text{J}/E_\text{C}=0.2\). The \(N\rightarrow \infty\) limit of \(\langle \hat{n} \rangle\) is plotted as a dotted blue curve. For \(\abs{n_\text{g}}\ll N\), the infinite-island expectation is recovered. For \(\abs{n_\text{g}}\gg N\), the staircase saturates at \(\langle\hat{n}\rangle\approx \pm N\). The corresponding (dimensionless) charge susceptibility \(d\langle\hat{n}\rangle/dn_\text{g}\) vs. offset charge \(n_\text{g}\) is plotted in (b). For \(E_\text{J}/E_\text{C}\ll 1\), the magnitude of the peaks occurring at \(n_\text{g}\in \{-N+1/2,...,N-1/2\}\) are given by Eq.~\eqref{eq:susceptibility} (solid black curve).}
    \label{fig:steps}
\end{figure}
\begin{align}\label{eq:first_order_freq}
    %\begin{aligned}
    \hbar \omega_\mathrm{q} &\approx \epsilon + {\bra{0}_b \hat{b} \delta\hat{H} \hat{b}^\dagger \ket{0}_b - \bra{0}_b \delta\hat{H} \ket{0}_b}\nonumber\\
    & \approx \epsilon + 2\mathrm{Re}\left[\bra{0}_b \hat{b}\commutator{E_\mathrm{C}\sqrt{N}\hat{p}'\delta \hat{S}_z}{\hat{b}^\dagger}\ket{0}_b\right]\nonumber\\
    & \approx \epsilon - \frac{E_\mathrm{C}}{2}\mathrm{Re}\left[\bra{0}_b \hat{b}\commutator{\hat{p}'\hat{a}^\dagger\hat{p}\hat{a}}{\hat{b}^\dagger}\ket{0}_b\right]\nonumber\\
    &\approx \sqrt{2E_\mathrm{C}E_\mathrm{J}}\left[1-\left(\frac{n_\mathrm{g}}{2N}\right)^2\right].
\end{align}
In the first line, we have introduced the vacuum of Bogoliubov quasiparticles  \(\ket{0}_b\) satisfying \(\hat{b} \ket{0}_b = 0\); in the second line, we have dropped the higher-order \(\delta\hat{S}_z^2\) term and introduced the shifted momentum \(\hat{p}' \equiv \hat{p}-n_\mathrm{g}/\sqrt{N}\); in the third line, we have truncated the Taylor expansion [Eq.~\eqref{eq:taylor}] at first order to yield \(\delta \hat{S}_z \approx -\hat{a}^\dagger\hat{p}\hat{a}/\sqrt{16N}\); and in the final line, we have expanded \(\hat{a}\) and \(\hat{a}^\dagger\) in terms of \(\hat{b}\) and \(\hat{b}^\dagger\), evaluated the vacuum expectation value, and approximated the result with the assumption \(1 \ll E_\mathrm{J}/E_\mathrm{C} \ll N^2\), which is easily satisfied by a typical transmon used in the field of quantum information. We note that the presence of \(n_\mathrm{g}\) in \(\delta \hat{H}\) renders higher-order corrections to Eq.~\eqref{eq:first_order_freq} increasingly important as \(|n_\mathrm{g}|\) increases.
 A numerical verification of this perturbative result [Eq.~\eqref{eq:first_order_freq}] is given in the Appendix. We conclude our analysis of the qubit frequency by emphasizing that Eq.~\eqref{eq:first_order_freq} may be tested by sweeping the gate voltage of a transmon and measuring the redshift in qubit frequency.

We now calculate the (dimensionless) charge susceptibility \(d\langle \hat{n} \rangle/dn_\text{g}\) by first computing the ground state \(\ket{\psi_0(n_\text{g})}\) of \(\hat{H}\) numerically and then computing the expected population imbalance \(\langle \hat{n} \rangle = \bra{\psi_0(n_\text{g})}\hat{n}\ket{\psi_0(n_\text{g})}\). Evaluating over a range of \(n_\text{g}\) and differentiating with a finite difference scheme yields the results in Fig.~\ref{fig:steps}. The numerical results showcase convergence to the predictions from the Mathieu equation~\cite{cottet2002implementation} for \(\abs{n_\mathrm{g}} \ll N\) and charge saturation for \(\abs{n_\mathrm{g}} \gg N\). Once again, representative analytical formulas are desirable. 

For the Cooper pair box regime \(E_\mathrm{J}/E_\mathrm{C}\ll 1\), we again employ degenerate perturbation theory at the degeneracy points \(n_\mathrm{g} \in \{-N+1/2,...,N-1/2\}\). Taking the expectation value of Eq.~\eqref{eq:two_level_number} in the ground state of Eq.~\eqref{eq:degenerate_H} yields \begin{equation}\label{eq:susceptibility}
    \frac{d\langle \hat{n}\rangle}{dn_\text{g}} \approx \frac{2NE_\text{C}}{E_\text{J}\sqrt{(1+2N)^2-4n_\mathrm{g}^2}},
\end{equation}
a good approximation at the aforementioned degeneracy points with \(E_\text{J}/E_\text{C} \ll 1\). In the infinite-island limit, the susceptibility [Eq.~\eqref{eq:susceptibility}] reduces to the well-known result \(\lim_{N\rightarrow \infty} (d\langle\hat{n}\rangle/dn_\mathrm{g}) = E_\mathrm{C}/E_\mathrm{J}\)~\cite{bouchiat1998quantum}.

As with the qubit frequency, an analytical expression for the charge susceptibility in the transmon regime \(E_\text{J}/E_\text{C} \gg 1 \) may be obtained via the Holstein-Primakoff transformation. To first order, the expected population imbalance reads
\begin{align}\label{eq:perturbative_pop_imbalance}
    \langle\hat{n}\rangle &\approx  \bra{0}_b\sqrt{N}\hat{p}\ket{0}_b + \bra{0}_b\delta\hat{S}_z\ket{0}_b + 2\mathrm{Re}\left[\bra{0}_b\sqrt{N}\hat{p}\ket{\delta\psi_0}\right]
\end{align}
with the first-order correction to the ground state \(\ket{\psi_0}\approx\ket{0}_b +\ket{\delta \psi_0}\) given by 
\begin{align}
    \ket{\delta\psi_0} &= \sum_{k>0}\frac{(\hat{b}^\dagger)^k\ket{0}_b\bra{0}_b\hat{b}^k\delta\hat{H}\ket{0}_b}{-k!k\epsilon}\nonumber \\
    &\approx  {\sum_{k>0}\frac{(\hat{b}^\dagger)^k\ket{0}_b\bra{0}_b \hat{b}^k \left(E_\mathrm{C}\sqrt{N}\hat{p}'\delta\hat{S}_z + \mathrm{H.c.}\right) \ket{0}_b}{-{k!}k\epsilon }}\nonumber\\
    &\approx \frac{E_\mathrm{C}}{4\epsilon}\sum_{k=1}^4\frac{(\hat{b}^\dagger)^k\ket{0}_b\bra{0}_b \hat{b}^k \left(\hat{p}'\hat{a}^\dagger\hat{p}\hat{a} + \mathrm{H.c.}\right) \ket{0}_b}{{k!}k}.
\end{align}
In the second line, we have dropped the higher-order \(\delta \hat{S}_z^2\) term, and in the third line, we have truncated the Taylor expansion [Eq.~\eqref{eq:taylor}] at first order. After applying the same truncation to the second term in Eq.~\eqref{eq:perturbative_pop_imbalance} and evaluating the relevant vacuum expectation values, we once again assume \(1 \ll E_\mathrm{J}/E_\mathrm{C} \ll N^2\) to extract the simple relation
\begin{equation}\label{eq:first_order_susceptibility}
    \frac{d\langle \hat{n}\rangle}{dn_\mathrm{g}} \approx 1 - \frac{3E_\mathrm{J}n_\mathrm{g}^2}{4E_\mathrm{C}N^4},
\end{equation}
which reduces to unity in the infinite-island limit, as expected.
A numerical verification of this result is given in the Appendix. This concludes our presentation of the main results. 

We have formulated a quantum theory of the JJ between finite superconducting islands, and we have used this theory to predict finite-island corrections to the qubit frequency and charge susceptibility. These predictions immediately motivate experiments in the \(n_\text{g} \sim N\) regime. While we have here focused on superconducting circuits, the present predictions are equally applicable to experiments involving trapped BECs. Finally, it is certainly possible and in fact experimentally observed (via parity switching~\cite{catelani2014parity}) that \(N\) may not remain a conserved quantity for all time. A modification to the present analysis capable of including such fluctuations is therefore desirable. The code used to generate all figures in this Letter, as well as to compute the analytical formulas for the qubit frequency and charge susceptibility in both the transmon and Cooper pair box regimes [Eqs.~\eqref{eq:qubit_frequency},~\eqref{eq:first_order_freq},~\eqref{eq:susceptibility}, and~\eqref{eq:first_order_susceptibility}] is openly available~\cite{repo}.

The authors are deeply grateful to Simon Geisert and Ioan Pop for helping us to ground this study in experimental reality, to Sean Crowe for calling our attention to~\cite{PhysRevA.55.4318} and encouraging us to explore multiple quantizations of the classical theory, to Nikhil Pimpalkhare for the contraction algorithm mentioned in the Appendix, and to Dung Pham, Zoe Zager, and Wentao Fan for their valuable feedback. Research supported by the U.S. Department of Energy, Office of Basic Energy Sciences, Division of Materials Sciences and Engineering under Award No. DESC0016011 (T. J. M., H. E. T.) and by a National Science Foundation Graduate Research Fellowship under Grant No. DGE-2039656 (T.J.M.). 
\bibliography{refs}% Produces the bibliography via BibTeX.
\appendix*
\section*{Appendix}
\begin{figure*}[tb]
  \centering
  \begin{minipage}[t]{0.48\textwidth}
    \centering
    \includegraphics[width=\linewidth]{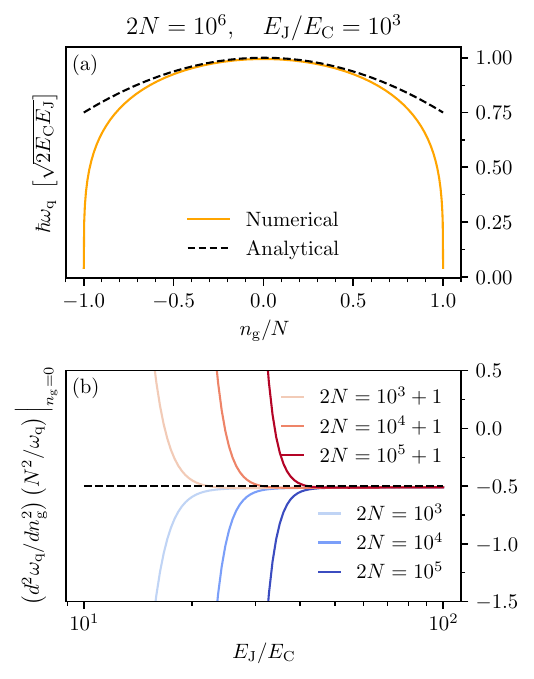}
    \caption{Panel (a) depicts the numerical verification of Eq.~\eqref{eq:first_order_freq} (dashed curve) by a representative example in the regime \(1\ll E_\mathrm{J}/E_\mathrm{C} \ll N^2\) (solid line) obtained via numerical diagonalization of Eq.~\eqref{eq:quantum_H}.  Panel (b) depicts the curvature of the charge dispersion at zero offset charge for a range of \(N\), each converging to the analytical result (dashed line) for sufficiently large \(E_\mathrm{J}/E_\mathrm{C}\). Parity switching of the curvature is emphasized.}   % → Figure 1
    \label{fig:transmon_frequency}
  \end{minipage}\hfill
  \begin{minipage}[t]{0.48\textwidth}
    \centering
    \includegraphics[width=\linewidth]{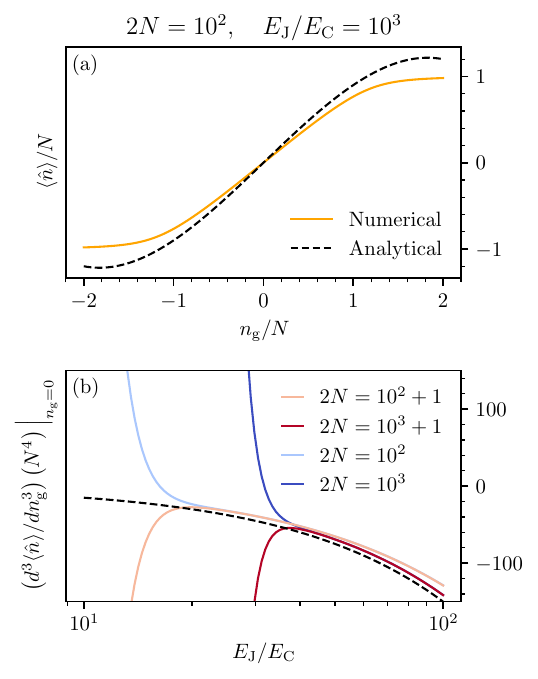}
    \caption{Panel (a) depicts the numerical verification of Eq.~\eqref{eq:first_order_susceptibility} (derivative of the dashed curve) by a representative example in the regime \(1\ll E_\mathrm{J}/E_\mathrm{C} \ll N^2\) (solid curve) obtained via numerical diagonalization of Eq.~\eqref{eq:quantum_H}. While better agreement can be found for larger values of \(N^2 E_\mathrm{C}/E_\mathrm{J}\), this example highlights the ability of perturbative corrections to capture the saturation occurring at \(\abs{n_\mathrm{g}}\gtrsim N\). Panel (b) depicts the curvature of the charge susceptibility at zero offset charge for a range of \(N\), each converging to the analytical result (dashed curve) for sufficiently large \(E_\mathrm{J}/E_\mathrm{C}\). Parity switching of the curvature is emphasized.}   % → Figure 2
    \label{fig:transmon_susceptibility}
  \end{minipage}
\end{figure*}

    Analytical formulas for the qubit frequency and charge susceptibility in the transmon regime \(1 \ll E_\mathrm{J}/E_\mathrm{C} \ll N^2\) are verified numerically in Figs.~\ref{fig:transmon_frequency} and \ref{fig:transmon_susceptibility}, respectively. We note that the requirement \(E_\mathrm{J}/E_\mathrm{C}\ll N^2\) may be lifted straightforwardly, though in this case, Eqs.~\eqref{eq:first_order_freq} and \eqref{eq:first_order_susceptibility} will become significantly more complicated. Evaluation of the relevant vacuum expectation values was performed using an in-house contraction algorithm.
    \noindent

\end{document}